\documentstyle[prl,aps,epsf,multicol]{revtex}

\tighten
\begin{document}
\draft

\title{Zero-Point Fluctuations and the Quenching of the
Persistent Current in Normal Metal Rings}

\author{Pascal Cedraschi, Vadim V.\ Ponomarenko, and Markus B\"uttiker}
\address{D\'epartement de Physique Th\'eorique,
Universit\'e de Gen\`eve,\\
24, quai Ernest Ansermet, CH-1211 Geneva 4, Switzerland}

\date{\today}

\maketitle

\begin{abstract}
The ground state of a phase-coherent mesoscopic system is sensitive to
its environment. We investigate the persistent current of a ring with
a quantum dot which is capacitively coupled to an external circuit
with a dissipative impedance. At zero temperature, zero-point quantum
fluctuations lead to a strong suppression of the persistent current
with decreasing external impedance.  We emphasize the role of
displacement currents in the dynamical fluctuations of the persistent
current and show that with decreasing external impedance the
fluctuations exceed the average persistent current.
\end{abstract}


\begin{multicols}{2}
\narrowtext

\begin{figure}
\centerline{\epsfysize=6cm\epsfbox{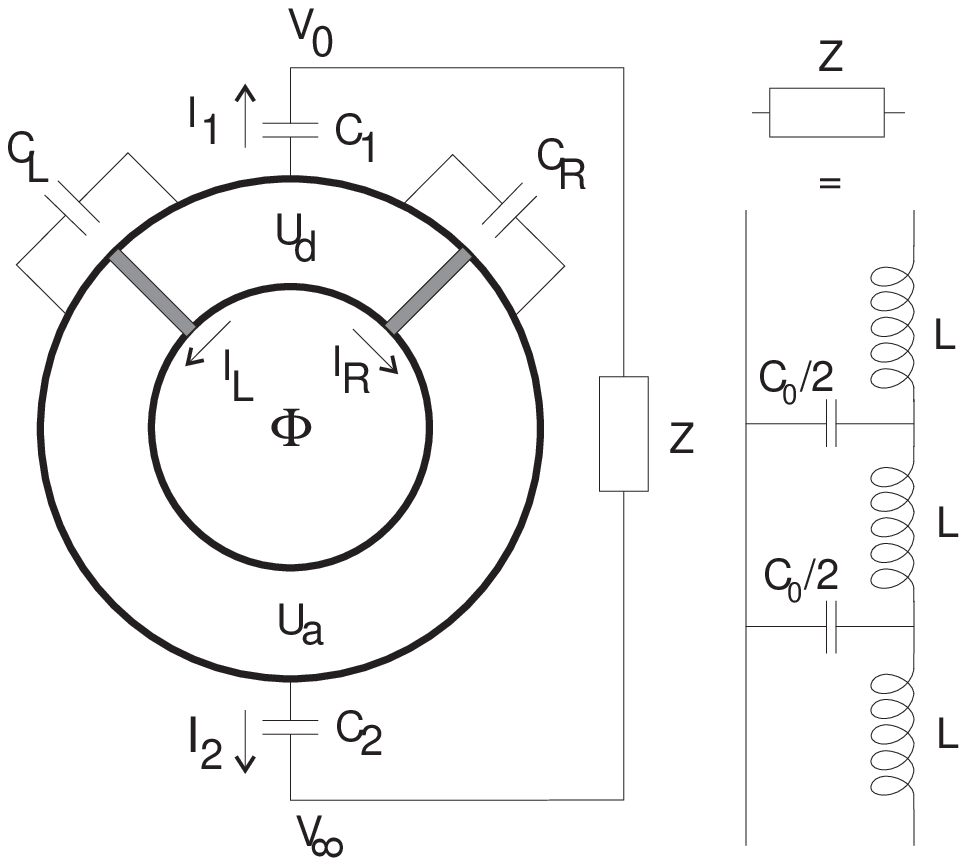}}
\vspace*{0.5cm}
\caption{\label{system}Ring with an in-line dot subject to a 
flux $\Phi$ and capacitively coupled to an external impedance $Z$.}
\end{figure}

The persistent current in a doubly connected conductor penetrated by
an Aharonov-Bohm flux is an indicator of quantum coherent electron
motion in the ground state of the system.  Only the charge carriers
whose wave functions are sufficiently extended to reach around the
loop can carry a persistent current \cite{BIL}.  In this letter, we
are interested in the effect of {\em zero-point fluctuations} of an
environment on the ground state of the system.  We consider a
mesoscopic ring with an in-line quantum dot, coupled capacitively to a
polarizable, dissipative environment, modeled by an impedance $Z$, see
Fig.~\ref{system}.  The fluctuations leading to exchange of charge
between the dot and the arm of the ring are capacitively coupled to
the charge fluctuations of the external circuit.  We investigate the
zero-temperature limit of this system in which the zero-point
fluctuations of the loop interact with the zero-point fluctuations of
the external circuit. We show that with decreasing external impedance
the persistent current in this system is strongly suppressed below its
value for infinite impedance. Thus the zero-temperature environment
can effectively destroy the coherence of the mesoscopic system. The
role of zero-temperature fluctuations and its effect on phase-coherent
transport properties is a subject of high current interest
\cite{mohant}. While the coherence properties of the ground state for
superconducting structures has been discussed \cite{hekk}, a discussion
for normal systems seems not to be available.

We investigate the properties of the ground state of the structure
depicted schematically in Fig.~\ref{system} with particular emphasis
on the influence of Coulomb interactions on the currents.  The Coulomb
interactions will be treated in a capacitive model which assumes that
well defined regions can be described with a single potential. In the
conductor of Fig.~\ref{system} the regions are the dot $(U_{d})$, the
arm $(U_{a})$ and the external circuit regions $(V_{0},V_{\infty})$
separated by capacitors $C_1$ and $C_2$ from the ring.  Our first task
is to derive expressions for the currents valid in the presence of
potential fluctuations.  For the dynamical fluctuations
\noindent
of interest here it is important to take into account that only {\em
total\/} currents, namely particle currents augmented by displacement
currents, are conserved quantities. It is the {\em total\/} currents
which are related to the magnetization and its fluctuations.  In terms
of the particle currents $I_L^p$ and $I_R^p$ through the left and
right tunnel barrier characterized by capacitances $C_{L}$ and $C_{R}$
the total currents are
\begin{eqnarray}
\label{il}
I_L&=&I_L^p+C_L{\partial\over\partial t}(U_d-U_a),\\
\label{ir}
I_R&=&I_R^p+C_R{\partial\over\partial t}(U_d-U_a),\\
\label{i1}
I_1&=&C_1{\partial\over\partial t}(U_d-V_0).
\end{eqnarray}
In $I_{1}$ the potential $V_{0}$ characterizes the external circuit.
Eqs.~(\ref{il}--\ref{i1}) are quantized by replacing $I_L^p$ and
$I_R^p$ by the corresponding particle current operators
$\hat{I}_{L,R}^p$ and by replacing time derivatives by commutators
with the Hamiltonian of the entire system $\partial A/\partial t \to
i/\hbar\,[\hat{H},\hat{A}]$.  Note that these currents satisfy
Kirchhoff's rule $I_L+I_R+I_1=0$, and thus are conserved, even when
quantized.  In the ground state the {\em average\/} currents through
the left and right junctions are equal to each other and both equal
(in magnitude) to the persistent current, $\langle\hat{I}_L\rangle =
-\langle\hat{I}_R\rangle = I = -c \, \partial F / \partial \Phi$,
where $F$ is the free energy.  However, the dynamical currents, in
particular, the {\em fluctuations\/} of $\hat{I}_L$ and $\hat{I}_R$
are usually different even in the ground state and depend explicitly
on the capacitances.

The circuit of Fig.~\ref{system} consists of two loops and can thus be
characterized by the {\em displacement} current through the external
loop, $\hat{I}_1 = -(\hat{I}_L + \hat{I}_R) = - \hat{I}_2$ and a
current {\em circulating} around the ring with the quantum dot,
$\hat{I}_c$.  Taking into account that the displacement current
$\hat{I}_1$ is divided over the two branches of the circuit according
to the capacitances $C_R$ and $C_L$, we find the circulating current
\begin{equation}\label{Ic}
\hat{I}_c = { C_R \hat{I}_L - C_L \hat{I}_R \over C }
\end{equation}
with $C \equiv C_L + C_R$.  Note that according to Eqs.~(\ref{il}) and
(\ref{ir}) this current depends on the capacitances and the 
particle portion of the total current. Such an expression for the
current, Eq.~(\ref{Ic}), is familiar from dynamical investigation of
resonant tunneling in double barrier structures, where it is known as
Ramo-Shockly theorem, but seems novel in the context of persistent
currents.

We model the finite impedance $Z$
in a Hamiltonian approach, with the help of a transmission line, see
Fig.~\ref{system}, with capacitance $C_{0}$ and inductance $L$.  For
convenience we take the capacitance to coincide with the effective
capacitance of the mesoscopic structure, $C_0^{-1}\equiv
C^{-1}+C_e^{-1}$ with $C_e^{-1} \equiv C_1^{-1} + C_2^{-1}$. The ohmic
resistance generated by the transmission line is $R\equiv
Z(\omega=0)=(2L/C_0)^{1/2}$ and thus can be varied by adjusting the
inductance.  There are other ways to model a resistance, and general
rules are given in Ref.~\cite{ingold:tunnel}.  We denote the charges
on the capacitors between the inductances by $Q_n$ and the potentials
by $V_n$, $n = 0, 1, 2, \ldots$.  Furthermore, we introduce the
generalized fluxes $\phi_n$ satisfying the equations
$d\phi_n/dt=V_n-V_\infty$.  The charge on the dot is $Q_d$, and the
charge on the capacitor $C_1$ is $Q_0$.  Then the Hamiltonian
including all electromagnetic interactions reads
\begin{eqnarray}
\label{HC}
H_C &=& {Q_d^2 \over 2C} + {Q_d Q_0 \over C} +{Q_0^2 \over 2C_0}
\nonumber \\ &+& \sum_{n=1}^\infty \left\{ {Q_n^2 \over C_0} +
{(\phi_n - \phi_{n-1})^2 \over 2L} \right\} .
\end{eqnarray}
Eq.~(\ref{HC}) is quantized by replacing the charge and the
generalized flux by operators obeying the commutation relations
$[\hat{\phi}_m, \hat{Q}_n] = i \hbar \delta_{mn}$.  The infinite sum
in Eq.~(\ref{HC}) is diagonalized by the transformation $Q_n =
\int_0^1 dx \, [ Q(x) + (C_0/C) \, e( N_+ + 1/2 ) ] \exp( -2\pi inx
)$, $\phi_n = \int_0^1 dx \, \phi(x) \exp( -2\pi inx )$, giving rise
to a bath of uncoupled harmonic oscillators
\begin{equation}
\label{HHO}
\hat{H}_{HO} = \int_0^1 dx \, \left\{ { \sin^2 \pi x \over L }
\hat{\phi}^2(x) + { \hat{Q}^2(x) \over 2 C_0 } \right\}.
\end{equation}
We observe that the spectrum of $\hat{H}_{HO}$ is dense, such that
there are no recurrence phenomena \cite{leggett:review}.  This
electromagnetic Hamiltonian, Eq.~(\ref{HC}), has to be augmented by
the Hamiltonian of the electronic degrees of freedom in the ring.  We
consider only spinless electrons in this letter and describe the
coupling of the dot and arm in the limit in which the tunneling
amplitudes through the left and the right junction are much smaller in
magnitude than the level spacing in the dot and the level spacing in
the arm. In this limit, discussed by Stafford and one of the
authors~\cite{buettiker:ringdot:prl}, one may consider hybridization
between the topmost occupied electron level in the arm and the lowest
unoccupied level in the dot, $\epsilon_{aM}$ and $\epsilon_{d(N+1)}$
only.  We denote the tunneling amplitudes between the levels
$\epsilon_{aM}$ and $\epsilon_{d(N+1)}$ by $t_L$ for tunneling through
the left junction and by $t_R$ for tunneling through the right one.
The total tunneling amplitude between the dot and the arm, is a
function of the Aharonov-Bohm flux $\Phi$ through the ring and is
given by $\hbar \Delta_0 / 2 \equiv ( t_L^2 + t_R^2 \pm 2 t_L t_R \cos
\varphi )^{1/2}$. The sign is positive if the number of electrons in
the ring is odd, and negative otherwise, and $\varphi \equiv 2\pi \Phi
/ \Phi_0$ is the Aharonov-Bohm flux in units of the flux quantum
$\Phi_{0} = hc/e$. Now, the Hamiltonian of the electronic degrees of
freedom reads $\hat{H}_e = [ ( \epsilon_{aM} - \epsilon_{d(N+1)} ) / 2
] \sigma_z - ( \hbar \Delta_0 / 2 ) \sigma_x$, and the charge on the
dot is $\hat{Q}_d = (e/2) \sigma_z + e( N - N_+ + 1/2 )$, where
$\sigma_z$ and $\sigma_x$ are Pauli matrices and $N_+$ is the number
of background charges on the dot.  Due to the interactions,
Eq.~(\ref{HC}), the {\em detuning} $\epsilon_{aM} - \epsilon_{d(N+1)}$
is changed into $\hbar \varepsilon \equiv \epsilon_{aM} -
\epsilon_{d(N+1)} + (e^2/C) ( N - N_+ + 1/2 )$.  The tunneling
described by the term in $\sigma_x$ is dressed by the interactions
with the dissipative circuit.  The complete Hamiltonian becomes
\begin{equation}
\label{HCL}
\hat{H} = { \hbar \varepsilon \over 2} \sigma_z + \hat{H}_{HO} - 
{ \hbar \Delta_0 \over 2 } \left( \sigma_+ e^{ -i \hat{\Omega} } +
\sigma_- e^{ i \hat{\Omega} } \right) ,
\end{equation}
where $\sigma_\pm = 1/2 ( \sigma_x \pm i \sigma_y )$ and $\hat{\Omega}
= ( e / \hbar ) ( C_0 / C ) \hat{\phi}_0$.  Eq.~(\ref{HCL}) is the
Caldeira-Leggett (CL) model \cite{leggett:review,caldeira:diss}.  To make
the connection to this model more explicit, we introduce the
RC-time $\tau_{RC} = RC$ and the
parameters $\omega_c = \pi \omega_0 / 2$ and the coupling strength
\begin{equation}
\alpha \equiv { \pi \over 2 } { e^2 / (2C) \over \hbar \omega_c }
{ 1 \over \omega_c \tau_{RC} }. 
\end{equation}
These parameters also 
determine the spectral density
$J(\omega) = 2\pi \hbar \alpha \omega \exp( -\omega / \omega_c )$.
The coupling between the ring and the external circuit is mediated by
the operator $\hat{\Omega}$.  We shall therefore investigate its
correlator, $\langle e^{ - i \hat{\Omega}(t) } e^{ i \hat{\Omega}(0) }
\rangle$, where $\hat{\Omega}(t) \equiv e^{ - i \hat{H} t / \hbar }
\hat{\Omega} e^{ i \hat{H} t / \hbar }$, in particular its long time
behavior.  In imaginary time, $\tau = it$, it reads
\begin{eqnarray}
\left\langle e^{ - i \hat{\Omega}(\tau) }
e^{ i \hat{\Omega}(0) } \right\rangle
&=& \exp \left[ -\int_0^\infty d\omega \,
{ J(\omega) \over \omega^2 } ( 1 - e^{ -\omega |\tau| } ) \right]
\nonumber \\
&=& \left( 1 + \omega_c |\tau| \right)^{-2\alpha} .
\end{eqnarray}
We note that $\alpha$ is
proportional to the Coulomb energy $e^2 / (2C)$, and inversely
proportional to the resistance $R$.  In the next section we discuss
the ground state of Eq.~(\ref{HCL}) for different values of the
coupling strength $\alpha$.

For $\alpha = 0$, the ring is not coupled to the bath of harmonic
oscillators, and the ground state is obtained by diagonalizing the $2
\times 2$-matrix $\hat{H}_{\alpha = 0} = ( \hbar \varepsilon / 2 )
\sigma_z - ( \hbar \Delta_0 / 2 ) \sigma_x$.  The persistent current
is $I = e/2 ( \varepsilon^2 + \Delta_0^2 )^{-1/2} \Delta_0 ( \partial
\Delta_0 / \partial \varphi )$ with $\hbar ( \partial \Delta_0 /
\partial \varphi ) = ( \mp t_L t_R \sin \varphi ) / \sqrt{ t_L^2 +
t_R^2 \pm 2 t_l t_R \cos \varphi } $ \cite{buettiker:ringdot:prl}.  It
is maximal at resonance, $\varepsilon = 0$, where it takes the value
$(e/2) \partial \Delta_0 / \partial \varphi$, and decreases like
$\varepsilon^{-1}$ far away from resonance.  The persistent current is
not only equal to the average of the circulating current, $I = \langle
\hat{I}_c \rangle$, but also equal (in magnitude) to the total as well
as to the particle currents through the left and the right junctions,
$I = \langle \hat{I}_L \rangle = \langle \hat{I}_L^p \rangle = -
\langle \hat{I}_R \rangle = - \langle \hat{I}_R^p \rangle$.  The
current through the ring vanishes on the average, $\langle \hat{I}_1
\rangle = 0$.  Thus the {\em average\/} currents are pure particle
currents.

In the {\em fluctuations\/} of the currents, however, displacement
currents play an important role.  In our model with only two relevant electron
levels, the fluctuations of the circulating current are given by the
general formula
\begin{eqnarray}
\label{deltaIC}
\Delta I_c{}^2 &=& \left( { e \over 2 } { \partial \Delta_0 \over
\partial \varphi } \right)^2 - I^2 \nonumber \\ &+& { e^2 \over
\hbar^2 } { \left[ C_R t_L^2 - C_L t_R^2 \pm ( C_R - C_L ) t_L t_R
\cos \varphi \right]^2 \over C^2 ( t_L^2 + t_R^2 \pm 2 t_L t_R \cos
\varphi ) } .
\end{eqnarray}
The first two terms are due to the particle currents through the
junctions whereas the last one is of electromagnetic origin.
Eq.~(\ref{deltaIC}) holds also when the ring is coupled to the
environment.  We point out that the fluctuations of the current are
minimal at resonance, where the current itself is maximal.  Note that
only the persistent current $I$ is changed by interaction with the
environment. This is a consequence of the effective two level
description of our system.  The fluctuations of the current through
the ring, $\hat{I}_1 = (2/L) \int_0^1 dx \, \sin^2 \pi x \,
\hat{\phi}(x)$ are entirely due to the displacement current, and for
$\alpha = 0$, they read
\begin{equation}
\label{deltaI1}
\Delta I_1{}^2 = {8 \over 3 \pi} e^2
{ \hbar \omega_c \over e^2/(2C) }
{ \omega_c \over \tau_{RC} } .
\end{equation}
They are inversely proportional to the charging energy as well as to
the resistance, thus they decrease with increasing charging energy,
and increase with increasing dissipation (decreasing $R$).  For
$\alpha \neq 0$, and far from resonance (even at resonance for $\alpha
> 1$), these fluctuations receive perturbative corrections in
$\Delta_0 / \omega_c$.  The dominating contribution to $\Delta
I_1{}^2$, however, comes from the high frequency oscillators, and thus
$\Delta I_1{}^2$ is not model independent.

We shall now determine the properties of the ground state at arbitrary
$\alpha > 0$.  It is useful to mention at this point that the CL
Hamiltonian, Eq.~(\ref{HCL}), may be mapped \cite{leggett:review} on
the Hamiltonian of the anisotropic Kondo model as well as on the
Hamiltonian of the resonant level model.  For a vanishing magnetic
field, corresponding to our model being at resonance, $\varepsilon =
0$, the Kondo model is characterized by two parameters, which may be
identified as $\alpha$ and $\Delta_0 / \omega_c$.  The half plane
spanned by $\Delta_0 / \omega_c \ge 0$ and $\alpha$ is divided up into
three regions by the separatrix equation $|1-\alpha| = \Delta_0 /
\omega_c$, where the parameters flow to different fixed points under
the action of the renormalization group \cite{anderson:kondo}.  We are
not interested in the regime of strong tunneling $\Delta_0 / \omega_c
> |1-\alpha|$.  Of the remaining two, the region $\alpha>1$, $\Delta_0
/ \omega_c < \alpha - 1$ corresponds to the ferromagnetic Kondo model
where $\Delta_0 / \omega_c$ renormalizes to zero, the region
$\alpha<1$, $\Delta_0 / \omega_c < 1 - \alpha$ corresponds to the
anti-ferromagnetic Kondo model that has been solved by Bethe ansatz
\cite{tsvelick:kondo}.  For large detuning $\varepsilon$, both cases
can be treated by a perturbative expansion in $\Delta_0$.

At zero temperature, to second order in $\Delta_0$, the free energy is
$F_0 + \delta F$ with $F_0 = - \hbar |\varepsilon| / 2$ and $\delta F
= $
\begin{equation}
\label{perturbativeF}
- { \hbar \omega_c \over 4 } \left( { \Delta_0 \over
\omega_c } \right)^2 e^{ | \varepsilon |/ \omega_c } \left| {
\varepsilon \over \omega_c } \right|^{ 2\alpha - 1 } \Gamma \left( 1 -
2\alpha, { | \varepsilon | \over \omega_c } \right),
\end{equation}
where $\Gamma(\zeta, x)$ is the incomplete gamma function.
Eq.~(\ref{perturbativeF}) is valid at any $\alpha$ up to second order
in $\Delta_0 / \omega_c$ and for large detuning $\varepsilon$.  In the
regime $\alpha - 1 \gg \Delta_0 / \omega_c$ (the ferromagnetic regime
of the Kondo model), Eq.~(\ref{perturbativeF}) is valid even for
arbitrary detuning $\varepsilon$.  It is well known
\cite{chakravarty:sb} that the CL model exhibits symmetry
breaking for $\alpha > 1$, namely $\langle \sigma_z \rangle$ exhibits
a finite jump as $\varepsilon$ crosses zero.  In perturbation theory
the width of this jump is $\langle \sigma_z \rangle_{\varepsilon = 0-}
- \langle \sigma_z \rangle_{\varepsilon = 0+} = 1 - (\Delta_0 /
\omega_c)^2 / (2 (2\alpha-1) (2\alpha-2))$.  In the persistent
current, it shows up as a cusp at resonance.

For $0 < \alpha < 1$, we use the Bethe ansatz
solution of the anti-ferromagnetic Kondo model, or rather of the
equivalent resonant level model \cite{ponomarenko:reslev}.  The low
energy properties of the problem depend on three energy scales, namely
the detuning $\varepsilon$, a cutoff $D$ and the ``Kondo temperature''
$T_K = \Delta_0 ( \Delta_0 / D )^{(\alpha/(1-\alpha))}$, which
is the only scale that depends on the magnetic flux $\varphi$.  The
persistent current is calculated from the known expression
\cite{ponomarenko:reslev} for $\langle \sigma_z \rangle = \partial F /
\partial (\hbar \varepsilon)$
\begin{eqnarray}
\left\langle \sigma_z \right\rangle 
= { i \over 4 \pi^{3/2} } 
&{}& \int_{-\infty}^\infty { dp \over p - i0 }
e^{ ip ( \ln z + b ) } \nonumber \\
&{}& \times { \Gamma( 1 + ip ) \Gamma( 1/2 - i(1-\alpha)p )
\over \Gamma( 1 + i\alpha p ) } ,
\end{eqnarray}
\begin{figure}
\centerline{\epsfysize6.5cm\epsfbox{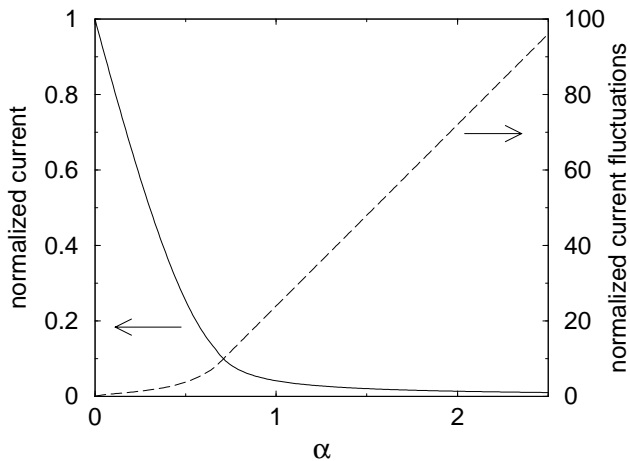}}
\caption{\label{kondopc} The persistent current at resonance (solid
line) in units of the current at $\alpha = 0$ and the normalized
fluctuations of the circulating current $\sqrt{\Delta I_c{}^2} / I$
(dashed line).  The parameters are $\omega_c = 25 \Delta_0$.}
\end{figure}
\noindent
where $z = ( \varepsilon / T_K )^{2(1-\alpha)}$ and $b = \alpha \ln
\alpha + (1-\alpha) \ln (1-\alpha)$.  This Bethe ansatz solution is
valid only for $1-\alpha \gg \Delta_0 / \omega_c$.  Now, we make use
of the fact that in terms of the above-mentioned energy scales, the
free energy reads $F = \hbar T_K f( \varepsilon / T_K, T_K / D )$, and
thus the persistent current $I = -(e / \hbar) (\partial F / \partial
\varphi)$ may be expressed in terms of $\langle \sigma_z \rangle =
\partial F / \partial (\hbar \varepsilon)$
\begin{equation}
\label{avI} 
I = -e {\partial T_K \over \partial \varphi} \left[
\int_0^y dx \, \langle \sigma_z \rangle (x)
- y \langle \sigma_z \rangle ( y ) \right] + I_0 .
\end{equation}
where $y \equiv {\varepsilon \over T_K}$.  
We point out that $T_K$ sets the scale for the transition from
resonant to perturbative behavior, i.e.\ for $\varepsilon \gg T_K$,
the expressions for the persistent current from Bethe ansatz and from
the perturbation theory must coincide up to terms of order $\Delta_0$.
This observation allows us to determine the cutoff $D$ in terms of
$\omega_c$,
\begin{equation}
\label{cutoff}
\left( { D \over \omega_c } \right)^{2\alpha}
= {2 \Gamma( 3/2 - \alpha ) e^{-b}
\over \sqrt{\pi} ( 1 - 2\alpha ) \Gamma( 1 - 2\alpha ) 
\Gamma( 1 - \alpha ) } ,
\end{equation}
as well as the integration constant $I_0$.
\begin{eqnarray}
\label{I0}
I_0 = &-& { e \over 2} 
{ \Gamma( 1 - { 1 \over 2 ( 1 - \alpha ) } )
e^{ - { b \over 2 ( 1 - \alpha ) } }
\over \sqrt{\pi} ( 1 - \alpha )
\Gamma( 1 - { \alpha \over 2 ( 1 - \alpha ) } ) }
\left( { \Delta_0 \over D } \right)^{ \alpha \over 1 - \alpha }
{ \partial \Delta_0 \over \partial \varphi } \nonumber \\
&+& { e \over 2} { 1 \over 1 - 2 \alpha }
{ \Delta_0 \over \omega_c } { \partial \Delta_0 \over \partial \varphi }.
\end{eqnarray}
The poles in the first term in Eq.~(\ref{I0}) appearing at $\alpha =
(2n+1)/(2n+2)$ for $n \ge 0$ are canceled by terms of higher order in
$\Delta_0$.  For $\alpha < 1/2$, the first term in Eq.~(\ref{I0}),
behaving like $( \Delta_0 / D )^{\alpha/(1-\alpha)}$ is dominating.
For $\alpha > 1/2$ the second one dominates and behaves as $\Delta_0 /
D $.  The power law for $\alpha > 1/2$ is thus the same as one obtains
from perturbation theory at $\alpha > 1$, using $I = -(e / \hbar) (
\partial F / \partial \varphi )$ with $F$ given by
Eq.~(\ref{perturbativeF}).  It is known that the lowest {\em excited}
state \cite{leggett:review} loses its phase coherence at $\alpha =
1/2$.  However, the persistent current remains differentiable in
$\varepsilon$ as $\alpha$ passes through $1/2$ and the ground state
retains some coherence even for large $\alpha$. As a function of
detuning the cusp at resonance shows up only at $\alpha > 1$.  The
poles at $\alpha = 1/2$ in the two terms in Eq.~(\ref{I0}) cancel each
other.  Together they give rise to a logarithmic persistent current
\begin{equation}
I_0\left( \alpha = { 1 \over 2 } \right)
\approx e { \partial \Delta_0 \over \partial \varphi }
{ \Delta_0 \over \omega_c } 
\left( \ln { \Delta_0 \over \omega_c } - 0.217 \ldots \right).
\end{equation}
The logarithmic term $\ln ( \Delta_0 / \omega_c )$ characterizes the
transition from power law to linear behavior.  The average circulating
current, Eq.~(\ref{avI}), is shown in Fig.~\ref{kondopc} as function
of $\alpha$ together with the mean squared fluctuations of the
circulating current obtained by using Eqs.~(\ref{avI}, \ref{I0}) and
Eq.~(\ref{deltaIC}).

Let us discuss some orders of magnitude.  Assume that the bosonic
excitations described by the bath of harmonic oscillators,
Eq.~(\ref{HHO}), are surface plasmons. Thus $\omega_c$ is
approximately the frequency of a surface plasmon $\omega_p$, which is
of the order of $\omega_p \approx 10^{14} s^{-1}$.  If we a
use a typical mesoscopic capacitance of $C \approx 10^{-16}F$, and $R
\approx 1 \Omega$, we obtain $\alpha \approx 1 (!)$. Clearly a large
range of coupling constants can be experimentally realized and thus 
permit the investigation of the effect of the zero-point fluctuations 
on the persistent current, and thus on the ground state, of a mesoscopic
system.

This work was supported by the Swiss National Science Foundation.

\end{multicols}

\end{document}